\documentclass[twoside,a4paper,11pt,reqno]{amsart}
\usepackage{amsmath,mathrsfs}
\usepackage{tabularx}%\usepackage{rotating}
\numberwithin{equation}{section}

\def\cal{\mathcal}
\newtheorem{theorem}{Theorem}[section]

\theoremstyle{definition}

\theoremstyle{remark}
\newtheorem{remark}[theorem]{Remark}
\numberwithin{equation}{section}
\begin{document}

\title[Relativistic Boltzmann Equation with Hard Interactions]
{Existence of Global Solution of the Cauchy Problem for the
Relativistic Boltzmann Equation with Hard Interactions}

\author{Zhenglu Jiang}
\address{Department of Mathematics, Zhongshan University,
Guangzhou 510275, P.~R.~China}
\email{mcsjzl@mail.sysu.edu.cn}
%\thanks{ZJ is supported by NSFC 10271121, 10511120278 and 10611120371,  and sponsored by
%SRF for ROCS, SEM}

\author{Lijun  Ma}
\address{ Department of Mathematics, Zhongshan University,
Guangzhou 510275, P.~R.~China}
\email{mljzsu@163.com}
%\thanks{}

%    General info

\date{\today.}



\begin{abstract}
By using the DiPerna and Lions techniques
for the nonrelativistic Boltzmann equation, it is shown that
there exists a global mild solution to the Cauchy problem for
the relativistic Boltzmann equation
with the assumptions of the relativistic scattering cross section including
some  relativistic  hard interactions and the initial data satifying
finite mass, ``inertia'', energy  and entropy.
\end{abstract}

\maketitle

\section{Introduction}\label{intro}
This paper is concerned with the study of the Cauchy problem for the relativistic Boltzmann equation
(or briefly, RBE) \cite{gv} of the form
\begin{equation}
\frac {\partial f}{\partial t}+\frac {{\bf p}}
{p_0}\frac{\partial f}{\partial{\bf x}}
=Q(f, f),   \label{rbe}
\end{equation}
where ${f=f(t, {\bf x}, {\bf p})}$ is a distribution function of a
one-particle relativistic gas without external forces, $t\in
(0,+\infty),$ ${\bf x}\in{\bf R}^3,$ ${\bf p}\in{\bf R}^3,$
$p_0=(1+|{\bf p}|^2)^{1/2}$ and $Q(f,f)$ is the relativistic
collision operator whose structure is described below. By using a
similar technique to one given in the nonrelativistic case,  this
paper is to deduce the existence of a global mild solution to RBE
(\ref{rbe}) given an initial condition $f|_{t=0}=f_0({\bf x},{\bf
p})$ in ${\bf R}^3\times{\bf R}^3$ which satisfies
\begin{equation}
f_0\mathop{\geq}\limits^{a.e.}0,
\iint_{{\bf R}^3\times{\bf R}^3}f_0(1+|{\bf x}|^2+p_0+|\ln f_0|)d^3{\bf x}d^3{\bf p}<\infty.
\label{rbec}
\end{equation}
In (\ref{rbec}), the second term of the integral can be regarded as a finite
initial ``inertia'' of the relativistic system while the fourth one represents
the Boltzmann entropy at an initial time.
The two other terms of the integral in (\ref{rbec}), from left to right,
 respectively, represent the mass and the energy
in the relativistic system at the initial time.  The finiteness of  all the integrals
states that the relativistic system has  finite mass,
``inertia'', energy and entropy
at the initial state.

If $\varphi\in {\cal D}({\bf R}^3),$ then $Q(\varphi,\varphi)$ is a function of
${\bf p}$ given by
\begin{equation}
Q(\varphi,\varphi)=\frac{1}{p_0}\int_{{\bf R}^3}\frac{d^3 {\bf p}_1}{p_{1 0}}
\int_{{S}^2}d\Omega [\varphi({\bf p}^\prime)\varphi({\bf p}_1^\prime)-\varphi({\bf p})\varphi({\bf p}_1)]B(g, \theta)
\label{rbek}
\end{equation}
where the different parts are explained below.
In (\ref{rbe}),
$Q(f,f)$ means $Q(f(t, {\bf x}, \cdot),f(t, {\bf x}, \cdot)).$

 ${\bf p}$ and ${\bf p}_1$ are dimensionless momenta
of two colliding particles immediately before collision while
${\bf p}^\prime$ and ${\bf p}_1^\prime$ are, respectively,
 dimensionless  momenta of the particles
corresponding to ${\bf p}$ and ${\bf p}_1$ immediately after collision;
$p_0=(1+|{\bf p}|^2)^{1/2}$ and $p_{10}=(1+|{\bf p}_1|^2)^{1/2}$
are, respectively,  the dimensionless energy
of the particles with the momenta ${\bf p}$ and ${\bf p}_1$
while $p_0^\prime=(1+|{\bf p}^\prime|^2)^{1/2}$ and
$p_{10}^\prime=(1+|{\bf p}_1^\prime|^2)^{1/2}$
are, respectively,  the dimensionless energy of  the particles with the momenta ${\bf p}^\prime$ and ${\bf p}_1^\prime$.
As is standard,  $\varphi_1=\varphi( {\bf p}_1)$
is denoted by $\varphi_1$, etc., and primes are used to represent
the results of collisions.
${\bf R}^3$ is a three-dimensional Euclidean space and $S^2$  a unit sphere surface.
$B(g, \theta)$ is given by $B(g, \theta)= gs^\frac{1}{2}\sigma(g, \theta)/2,$
where $\sigma(g, \theta)$ is a scattering cross section,
$s=|p_{10}+p_0|^2-|{\bf p}_1+{\bf p}|^2,$
$g=\sqrt{|{\bf p}_1-{\bf p}|^2-|p_{10}-p_0|^2}/2,$ $ \theta$ is the  scattering angle
defined in $[0, \pi]$ by $\cos\theta
=1-2[(p_0-p_{10})(p_0-p_0^{\prime})-({\bf p}-{\bf p}_1)({\bf p}-{\bf p}^{\prime})]/(4-s).$
Obviously, $s=4+4g^2$.
$d\Omega=\sin{\theta}d{\theta}d{\psi}, $ $0{\leq}\theta{\leq}\pi,$ $
0{\leq}\psi{\leq}2\pi. $

The DiPerna and Lions techniques for the nonrelativistic Boltzmann
equation were first applied by Dudy\'{n}ski and Ekiel-Je\.{z}ewska
\cite{de92} to prove global existence of solutions to the Cauchy
problem for RBE with the assumptions of the relativistic scattering
cross section excluding the relativistic hard interactions and
 the initial data
satisfying finite mass, energy and entropy.
Unlike in the nonrelativistic case, the causality of solutions
to RBE is used by Dudy\'{n}ski and Ekiel-Je\.{z}ewska into their proof and so
the relativistic initial data is not required to have finite  ``inertia''.
Their results are correct but  the relativistic scattering cross section
does not include the cases about the relativistic hard interactions.
After that, a different device was also given in \cite{j98a}
to show global existence of solutions to the large-data Cauchy problem
for RBE with some relativistic hard interactions;
in this proof, the property of the causality of RBE is not used directly in
solving the Cauchy problem but it is assumed that
the initial data satisfies
\begin{equation}
f_0\mathop{\geq}\limits^{a.e.}0,
\iint_{{\bf R}^3\times{\bf R}^3}f_0(1+p_0|{\bf x}|^2+p_0+|\ln f_0|)d^3{\bf x}d^3{\bf p}<\infty,
\label{rbec2}
\end{equation}
i.e., finite mass, ``inertia'', energy
and entropy.
The objective of this paper is to show that there exists a global
mild solutions to the large-data Cauchy problem
for RBE with some relativistic hard interactions under the condition of the initial data
 $f_0$ satisfying (\ref{rbec}), that is,
\begin{theorem}\label{th1}
Let $B(g, \theta)$ be the relativistic collision kernel of RBE (\ref{rbe}), defined above,
and $B_R$ a ball with a center at the origin and a radius $R,$
$A(g)=\int_{S^2}d\Omega B(g, \theta).$
Assume that
\begin{equation}
B(g, \theta)\geq 0 ~\hbox{ a.e. in } [0,+\infty)\times S^2,
B(g, \theta)\in L_{loc}^1({\bf R}^3\times S^2),
\label{rbeb1}
\end{equation}
\begin{equation}
\frac{1}{p_0^2}\int_{B_R}\frac{d^3{\bf p}_1}{p_{10}}A(g)
{\rightarrow}0~ \hbox{ as  } |{\bf p}|{\rightarrow}+\infty,  ~\hbox{ for }{\forall}
R{\in}(0,+\infty).
\label{rbeb2}
\end{equation}
Then RBE (\ref{rbe}) has a mild
or equivalently a renormalized solution $f$
through initial data $f_0$ with (\ref{rbec}),
satisfying the following properties
\begin{equation}f\in C([0,+\infty);L^1({\bf R}^3{\times}{\bf R}^3)),
\label{sc1}\end{equation}
\begin{equation}L(f)\in L^\infty([0,+\infty);L^1({\bf R}^3{\times}B_R))
~\hbox{ for all }~  {\forall}R{\in}(0,+\infty),
\label{sc2}\end{equation}
\begin{equation}\frac{Q^+(f,f)}{1+f}\in L^1([0,T];L^1({\bf R}^3{\times}B_R))
~\hbox{ for }~  {\forall}R{\in}(0,+\infty)\hbox{ and }{\forall} T{\in}[0,+\infty),
\label{sc3}\end{equation}
\begin{equation}
\sup\limits_{0\leq t\leq T}\iint_{{\bf R}^3{\times}{\bf R}^3}f(1+
|{\bf x}|^2+p_0
+|{\ln}f|)d^3{\bf x}d^3{\bf p}<C_T,
\label{sc4}\end{equation}
where $C_T$ is a positive constant only dependent of $f_0$ and $T$
for ${\forall}T\in [0,+\infty).$
\end{theorem}
Obviously, the assumption of (\ref{rbec})
is weaker than that of (\ref{rbec2}) and so this result is better
than that obtained in \cite{j98a}.  Also, our proof is simpler than that given in \cite{j98a}.

It is clear that the condition (\ref{rbeb1}) is equivalent to the following one:
\begin{equation}
\sigma(g, \theta)\mathop{\geq}\limits^{a.e.}0\hbox{ in }
[0,+\infty)\times S^2, g(1+g^2)^{1/2}\sigma(g, \theta)\in
L_{loc}^1([0,+\infty)\times S^2), \label{rbeb12}
\end{equation}
 which was first defined by Jiang \cite{j98a}.
The assumption (\ref{rbeb2}) was originally introduced
by Jiang  \cite{j97a}\cite{j99}.
Obviously, the relativistic assumptions  (\ref{rbeb1}) and  (\ref{rbeb2}) are similar
to the following nonrelativistic ones adopted by DiPerna and Lions \cite{dl}:
\begin{equation}
B({\bf z}, \omega)\geq 0 ~\hbox{ a.e. in } {\bf R}^{N}\times S^{N-1},
B({\bf z}, \omega)\in L_{loc}^1({\bf R}^{N}\times S^{N-1}),
\label{beb1}
\end{equation}
\begin{equation}
\frac{1}{1+|{\bf \xi}|^2}\int_{|{\bf z}-{\bf \xi}|\leq R}d{\bf z}\tilde{A}({\bf z})
{\rightarrow}0~ \hbox{ as  } |{\bf \xi}|{\rightarrow}+\infty,  ~\hbox{ for }{\forall}
R{\in}(0,+\infty),
\label{beb2}
\end{equation}
where $B({\bf z}, \omega)$ is a function of $|{\bf z}|,$ $|({\bf z},\omega)|$ only,
$\tilde{A}({\bf z})=\int_{S^{N-1}}d\omega B({\bf z}, \omega).$
It is also easy to see that the condition (\ref{rbeb2}) includes some
relativistic hard interactions defined as $\int_{S^{2}}d\Omega B(g, \theta)\geq Cg^2,$
where $C$ is a positive constant (see \cite{de88}).
But it was assumed by Dudy\'{n}ski  and
Ekiel-Je\.{z}ewska (see \cite{de92}) that $B(g, \theta)$ satisfies (\ref{rbeb1}) and the following condition:
\begin{equation}
\frac{1}{p_0}\int_{B_R}\frac{d^3{\bf p}_1}{p_{10}}A(g)
{\rightarrow}0~ \hbox{ as  } |{\bf p}|{\rightarrow}+\infty,  ~\hbox{ for }{\forall}
R{\in}(0,+\infty),
\label{rbeb2de}
\end{equation}
where $B_R$ and $A(g)$
are the same as (\ref{rbeb2});  it has been claimed in \cite{de92} that
their assumptions of $B(g, \theta)$ exclude the relativistic hard interactions.
It follows that  (\ref{rbeb2de}) is more restrictive than (\ref{rbeb2}).

Apart from those mentioned above, there are many other outstanding results relevant to
the study of the Cauchy problem for RBE,
e.g.,  the works of Andr\'easson, Bancel, Bichteler, Cercignani, Glassey, Kremer, Strauss and so on
(see \cite{ka}\cite{bd}\cite{b}\cite{gs91}\cite{gs92}\cite{gs}).
It is worth mentioning that the recent work of Glassey \cite{g06} not only
introduces many relevant books and papers but also is a breakthrough.
These are very helpful to our further studying this problem.

\section{Conservation Laws and Entropy}
\label{cons}
As in the nonrelativistic case, the structure of the relativistic collision operator
maintains not only the conversation of mass, momenta
and energy in the relativistic system,  but also
the property that the entropy of the system does not decrease.

For convenience, let us first introduce the following notations:
\begin{equation}
Q^+(\varphi,\varphi)=\frac{1}{p_0}\int_{{\bf R}^3}\frac{d^3 {\bf p}_1}{p_{1 0}}
\int_{{S}^2}d\Omega \varphi({\bf p}^\prime)\varphi({\bf p}_1^\prime)B(g, \theta),
\label{rbekp}
\end{equation}
\begin{equation}
L(\varphi)=\frac{1}{p_0}\int_{{\bf R}^3}\frac{d^3 {\bf p}_1}{p_{1 0}}
\int_{{S}^2}d\Omega \varphi({\bf p}_1)B(g, \theta),
\label{rbekl}
\end{equation}
\begin{equation}
Q^-(\varphi,\varphi)=\varphi({\bf p})L(\varphi).
\label{rbekm}
\end{equation}
Obviously,
\begin{equation}
Q(\varphi,\varphi)=Q^+(\varphi,\varphi)-Q^-(\varphi,\varphi).
\label{rbekpm}
\end{equation}
Finally, it can be known that (\ref{rbek}) has the following equivalent form
\begin{equation}
Q(f, f)=\frac {1}{2}\mathop{\int\int\int}
\limits_{{\bf R}^3{\times}{\bf R}^3{\times}{\bf R}^3}
\frac{W({\bf p}, {\bf p}_1;{\bf p}^\prime, {\bf p}_1^\prime)}
{p_0p_{10}p_0^{\prime}p_{10}^\prime}
[f^{\prime}f_1^\prime-ff_1]d^3{\bf p}_1d^3{\bf p}^{\prime}d^3{\bf p}_1^\prime,
\label{rbektr}
\end{equation}
where
\begin{equation}
 W({\bf p}, {\bf p}_1;{\bf p}^\prime, {\bf p}_1^\prime)
=s\sigma(g, \theta){\delta}^{(3)}({\bf p}+{\bf p}_1-{\bf p}^\prime-{\bf p}_1^\prime)
\delta(p_0+p_{10}-p_0^\prime-p_{10}^\prime),
\label{rbetr}
\end{equation}
which is called transition rate for RBE (see \cite{gv}). Note that
energy and momenta of  two colliding particles conserve before and
after  collision. It can be then shown from (\ref{rbetr}) that the
transition rate for RBE satisfies
\begin{equation}
W({\bf p}, {\bf p}_1;{\bf p}^\prime, {\bf p}_1^\prime)
=W({\bf p}^\prime, {\bf p}_1^\prime;{\bf p}, {\bf p}_1)
=W({\bf p}_1, {\bf p};{\bf p}_1^\prime, {\bf p}^\prime)
=W({\bf p}_1^\prime, {\bf p}^\prime;{\bf p}_1, {\bf p}).
\label{rbetrp}
\end{equation}
By using (\ref{rbek}), (\ref{rbektr}) and (\ref{rbetrp}),
it can be easily proved that
\begin{equation}
\int_{{\bf R}^3}\psi({\bf p})Q(\varphi, \varphi)d^3{\bf p}
=\frac{1}{4}\iint_{{\bf R}^3\times{\bf R}^3}\frac{d^3 {\bf p}_1}{p_0p_{1 0}}
\int_{{S}^2}d\Omega B(g, \theta)
[\varphi^{\prime}\varphi_1^\prime-\varphi\varphi_1]
\nonumber
\end{equation}
\begin{equation}
\cdot[\psi({\bf p})+\psi({\bf p}_1)-\psi({\bf p}^\prime)-\psi(
{\bf p}_1^\prime)]
\label{rbekeq}
\end{equation}
where $ Q^\pm(\varphi, \varphi)\psi({\bf p})\in{L}^1({\bf R}^3)$
 for any given $\psi(
{\bf p})\in{L}^\infty({\bf R}^3) $ and
$\varphi({\bf p})\in{L}^1({\bf R}^3). $
It follows from (\ref{rbekeq}) that $\int_{{\bf R}^3}
{\bar \psi}Q(f, f)d^3{\bf p}=0$ if
$f=f(t,{\bf x}, {\bf p})$ is a distributional solution to RBE (\ref{rbe})
such that $\int_{{\bf R}^3}{\bar \psi}Q(f, f)d^3{\bf p}<+\infty$ for almost all $t$ and ${\bf x}$ and
${\bar \psi}={\bar b_0}+{\bf b}
{\bf p}+c_0p_0, $ where
 ${\bar b_0}\in{\bf R},
{\bf b}\in{\bf R}^3, c_0\in{\bf R}.$
Furthermore,  it is at least formally found that
$\iint_{{\bf R}^3\times{\bf R}^3}{\bar \psi}fd^3{\bf x}d^3{\bf p}$
 is independent of $t$  for any distributional solution $f$ to RBE (\ref{rbe}).
This yields the conservation of mass, momentum and energy of the
relativistic system.

Fortunately, the desired estimate of  the ``inertia''
$\iint_{{\bf R}^3{\times}{\bf R}^3}
f|{\bf x}|^2d^3{\bf x}d^3{\bf p}$ under the assumption of  (\ref{rbec})
can be also made successfully (see \cite{ck}).
This estimate is different from that of  the ``inertia'' $\iint_{{\bf R}^3{\times}{\bf R}^3}
fp_0|{\bf x}|^2d^3{\bf x}d^3{\bf p}$ defined by Jiang \cite{j98a}
under the assumption (\ref{rbec2}).
To show this estimate, it requires
the following identity
\begin{equation}
\frac{d}{dt}\iint_{{\bf R}^3{\times}{\bf R}^3}
f|{\bf x}|^2d^3{\bf x}d^3{\bf p}
=2\iint_{{\bf R}^3{\times}{\bf R}^3}
f({\bf x}{\bf p}/p_0)d^3{\bf x}d^3{\bf p}
\label{dinvariant2o}
\end{equation}
derived by multiplying RBE (\ref{rbe}) by $|{\bf x}|^2$
and integrating by parts over ${\bf x}$ and ${\bf p}.$
It follows from (\ref{dinvariant2o}) that
\begin{equation}
\frac{d}{dt}\iint_{{\bf R}^3{\times}{\bf R}^3}
f|{\bf x}|^2d^3{\bf x}d^3{\bf p}
\leq \iint_{{\bf R}^3{\times}{\bf R}^3}
f|{\bf x}|^2d^3{\bf x}d^3{\bf p}+\iint_{{\bf R}^3{\times}{\bf R}^3}
fd^3{\bf x}d^3{\bf p}.
\label{dinvariant2}
\end{equation}
This yields the following inequality
\begin{equation}
\sup\limits_{0\leq t\leq T}\iint_{{\bf R}^3{\times}{\bf R}^3}
f|{\bf x}|^2d^3{\bf x}d^3{\bf p}
\leq e^T \iint_{{\bf R}^3{\times}{\bf R}^3}
f_0(1+|{\bf x}|^2)d^3{\bf x}d^3{\bf p}
\label{dinvariant3}
\end{equation}
for any given $T>0$ by multiplying the two sides of (\ref{dinvariant2}) by $e^{-t}$
and  using the conservation of the mass of the  relativistic system.
The inequality given by (\ref{dinvariant3}) illustrates
that the relativistic ``inertia'' of $f|{\bf x}|^2$ over all the space and
momentum variables in any given time interval is controlled  by both mass and ``inertia''
at the initial state of the relativistic system.

Next, let us show the property that
the entropy is always a nondecreasing function of $t$
in the relativistic system.  To do this,
we first deduce at least formally the following relativistic entropy identity
\begin{equation}
\frac{d}{dt}\iint_{{\bf R}^3\times{\bf R}^3}f\ln fd^3{\bf x}d^3{\bf p}
+\frac{1}{4p_0}\iint_{{\bf R}^3\times{\bf R}^3}\frac{d^3 {\bf p}_1}{p_{1 0}}
\int_{{S}^2}d\Omega B(g, \theta)
\nonumber
\end{equation}
\begin{equation}
\cdot[f^{\prime}f_1^\prime-ff_1]\ln\left(\frac{f^{\prime}f_1^\prime}{ff_1}\right)=0
\label{rbeei}
\end{equation}
by multiplying RBE (\ref{rbe}) by $1+\ln f,$
integrating over ${\bf x}$ and ${\bf p}$
and using (\ref{rbekeq}).  In general,
$\iint_{{\bf R}^3\times{\bf R}^3}f\ln fd^3{\bf x}d^3{\bf p}$ is denoted by $H(t)$
and called H-function.
Boltzmann's entropy is usually defined by $-H(t).$
The second term in (\ref{rbeei}) is nonnegative and so
 $H(t)$ is a nonincreasing function of $t.$
This means that
the entropy of the relativistic system does not decrease.
This property allows the desired estimate of the relativistic entropy
to be derived from the Cauchy problem for RBE (\ref{rbe}).

The entropy can be controlled by the integral
$\iint_{{\bf R}^3\times{\bf R}^3}f|\ln f|d^3{\bf x}d^3{\bf p}$
for any nonnegative solution $f$ to RBE (\ref{rbe})
 and so  it is natural to make the considered estimate
of the integral instead of the entropy.
  Note that
\begin{equation}
\iint_{{\bf R}^3\times{\bf R}^3}f|\ln f|d^3{\bf x}d^3{\bf p}=\iint_{{\bf R}^3\times{\bf R}^3}f\ln fd^3{\bf x}d^3{\bf p}
+2\iint_{f\leq 1}f|\ln f|d^3{\bf x}d^3{\bf p}
\nonumber
\end{equation}
\begin{equation}
\hspace*{2cm}\leq \iint_{{\bf R}^3\times{\bf R}^3}f\ln fd^3{\bf x}d^3{\bf p}
+2\iint_{{\bf R}^3\times{\bf R}^3}f(|{\bf x}|^2+p_0)d^3{\bf x}d^3{\bf p}
\nonumber
\end{equation}
\begin{equation}
\hspace*{2cm}+2\iint_{f\leq \exp(-|{\bf x}|^2-p_0)}f\ln(1/f)d^3{\bf x}d^3{\bf p}
\nonumber
\end{equation}
\begin{equation}
\hspace*{1cm}\leq \iint_{{\bf R}^3\times{\bf R}^3}f\ln fd^3{\bf x}d^3{\bf p}
+2\iint_{{\bf R}^3\times{\bf R}^3}f(|{\bf x}|^2+p_0)d^3{\bf x}d^3{\bf p}
+C_1
\label{enineq}
\end{equation}
where $C_1$ is some positive constant independent of $f.$
By  using (\ref{rbeei}) and (\ref{enineq}), it can be then deduced that
\begin{eqnarray}
\sup\limits_{0\leq t\leq T}\left[\iint_{{\bf R}^3\times{\bf R}^3}f|\ln f|d^3{\bf x}d^3{\bf p}\right] \hspace*{3cm}\nonumber\\
\leq \iint_{{\bf R}^3\times{\bf R}^3}f_0[2e^T(|{\bf x}|^2+1)+2p_0+|\ln f_0|]d^3{\bf x}d^3{\bf p}+C_1.
\label{rbeest}
\end{eqnarray}

\section{Proof of Theorem \ref{th1}}
\label{pt}
In this section we show our theorem given in Section \ref{intro} by use of the DiPerna and Lions global existence proof with
only minor adjustments.

Let us first consider a similar approximation scheme
to that given by DiPerna and Lions \cite{dl} in the nonrelativistic case.
Put $f_0^n=\min(f_01_{|{\bf x}|^2+
|{\bf p}|^2{\leq}n}, n)+
\frac{1}{n}e^{-(|{\bf x}|^2+p_0)}$ and  $B_n(g, \theta)=gs^{\frac{1}{2}}\sigma_n(g, \theta)$
where  $\sigma_n(g, \theta)
=\sigma(g, \theta)1_{\sigma(g, \theta){\leq}n}(g, \theta)
1_{g{\geq}\frac{1}{n}}(g)1_{\sin\theta{\geq}
\frac{1}{n}}(\theta)1_{p_0+p_{10}\leq{n}}$
for $n=1, 2,$ $3, \cdots,$  and  through this paper, all the functions
denoted by $1$ with subscript expressions, such as $1_{|{\bf x}|^2+
|{\bf p}|^2{\leq}n},$ represent characteristic functions.
It follows that
\begin{equation}
\iint_{{\bf R}^3\times{\bf R}^3}d^3{\bf x}d^3{\bf p}
|f_0-f_0^n|(1+|{\bf x}|^2+p_0)\to 0 \hbox{ as } n\to +\infty,
\end{equation}
\begin{equation}
\iint_{{\bf R}^3\times{\bf R}^3}d^3{\bf x}d^3{\bf p}
f_0^n|\ln f_0^n|\leq C,
\end{equation}
where $C$ is a positive constant independent of $n.$
It is also found from (\ref{rbeb1}) that $B_n(g, \theta)\in L^\infty\cap L^1({\bf R}^3;L^1(S^2))$ and
\begin{equation}
\iint_{B_R\times S^2}d^3 {\bf p}d\Omega|B_n(g, \theta)-B(g, \theta)|\to 0
\label{arbekc}
\end{equation}
uniformly in $\{{\bf p}_1: |{\bf p}_1|\leq k\}$ as $n\to +\infty$ for $\forall R,k\in (0,+\infty).$

Then the collision kernel of RBE (\ref{rbe})
is replaced by $B_n(g, \theta)$
to solve the following Cauchy problem
\begin{equation}
%\left.\begin{array}{c}
\frac {\partial f^n}{\partial t}+\frac {{\bf p}}
{p_0}\frac {\partial f^n}{\partial{\bf x}}
=\tilde{Q}_n(f^n, f^n),
%\\
f^n|_{t=0}=f_0^n.
%\end{array}\right\}.
\label{arbe}
\end{equation}
Here and below, $\tilde{Q}_n$ is defined by
$\tilde{Q}_n(\varphi, \varphi)
=(1+\frac{1}{n}\int_{{\bf R}^3}|\varphi | d^3{\bf p})^{-1}Q_n(\varphi,\varphi)$ and
\begin{equation}
Q_n(\varphi,\varphi)=\frac{1}{p_0}\int_{{\bf R}^3}\frac{d^3 {\bf p}_1}{p_{1 0}}
\int_{{S}^2}d\Omega
[\varphi({\bf p}^\prime)\varphi({\bf p}_1^\prime)-\varphi({\bf p})\varphi({\bf p}_1)]
B_n(g, \theta).
\label{arbek}
\end{equation}
It follows from (\ref{arbek}) that for all $\varphi,\psi \in L^\infty([0, +\infty)\times{\bf R}^3{\times}{\bf R}^3)
 \cap L^1({\bf R}^3{\times}{\bf R}^3)),$
\begin{equation}
||\tilde{Q}_n(\varphi, \varphi)||_{L^\infty([0, +\infty)\times{\bf R}^3{\times}{\bf R}^3)}
\leq C_n||\varphi ||_{L^\infty([0, +\infty)\times{\bf R}^3{\times}{\bf R}^3)},
\label{arbek1}
\end{equation}
\begin{equation}
||\tilde{Q}_n(\varphi, \varphi)||_{L^1({\bf R}^3{\times}{\bf R}^3)}
\leq C_n||\varphi ||_{L^1({\bf R}^3{\times}{\bf R}^3)},
\label{arbek2}
\end{equation}
\begin{equation}
||\tilde{Q}_n(\varphi, \varphi)-\tilde{Q}_n(\psi, \psi)||_{L^1({\bf R}^3{\times}{\bf R}^3)}
\leq C_n||\varphi- \psi ||_{L^1({\bf R}^3{\times}{\bf R}^3)},
\label{arbek3}
\end{equation}
here and below everywhere, $C_n$ is a nonnegative constant independent of $\varphi$ and $\psi.$

Existence and uniqueness of distributional solutions
to the Cauchy problem given by (\ref{arbe})
can be below established.
The first step to show this is to construct
a set ${\cal F}$ such that
each $\varphi\equiv \varphi(t, {\bf x}, {\bf p}) $
belongs to ${\cal F}$ if and only if $\varphi$ is
a measurable function defined in $[0,+\infty){\times}{\bf R}^3{\times}{\bf R}^3 $
and satisfies
$$\sup\limits_{
(t, {\bf x}, {\bf p})
\in[0, +\infty)\times{\bf R}^3{\times}{\bf R}^3}e^{-C_nt}
|\varphi(t, {\bf x}, {\bf p})|
{\leq}||f_0^n||_{L^\infty({\bf R}^3{\times}
{\bf R}^3)}, \leqno(i)$$
$$\sup\limits_{0\leq{t}<+\infty}e^{-C_nt}\int\int_{{\bf R}^3\times{\bf R}^3}
|\varphi |d^3{\bf x}d^3{\bf p}{\leq}||f_0^n||_{L^1({\bf R}^3{\times}
{\bf R}^3)}. \leqno(ii)$$
The second is to define
a mapping on ${\cal F} $ as follows:
for each $\varphi\in{\cal F}, $
\begin{equation}
J_n(\varphi)(t, {\bf x}, {\bf p})
=f_0^n({\bf x}-t\frac{{\bf p}}{p_0}, {\bf p})
+\int_0^t\tilde{Q}_n(\varphi, \varphi)
(\sigma, {\bf x}
+(\sigma-t)\frac{{\bf p}}{p_0},
{\bf p})d\sigma.
\label{conop}
\end{equation}
Then, by (\ref{arbek1}),  (\ref{arbek2}) and (\ref{arbek3}),
it can be proved that $J_n({\cal F})\subseteq{\cal F}$ and that
 $J_n$ is contractive, i.e., for every $g, h\in{\cal F}, $
\begin{equation}
\sup\limits_{0{\leq}t<+\infty}e^{-2C_nt}\int\int_{{\bf R}^3\times
{\bf R}^3}|J_n(g)(t, {\bf x}, {\bf p})
-J_n(h)(t, {\bf x}, {\bf p})|d^3{\bf x}d^3{\bf p}
\nonumber
\end{equation}
\begin{equation}
{\leq}\frac{1}{2}\sup\limits_{0{\leq}t<+\infty}e^{-2C_nt}
\int\int_{{\bf R}^3\times{\bf R}^3}|g
(t, {\bf x}, {\bf p})
-h(t, {\bf x}, {\bf p})|d^3{\bf x}d^3{\bf p}.
\label{contractive}
\end{equation}
Therefore there exists a unique element $f^n\equiv f^n(t, {\bf x}, {\bf p})
\in {\cal F}
$ such that for almost every $
({\bf x}, {\bf p})\in
{\bf R}^3{\times}{\bf R}^3, $
\begin{equation}
(f^n)^\#(t, {\bf x}, {\bf p})
=J_n(f^n)^\#(t, {\bf x}, {\bf p})
~~\hbox{ for }~ \forall t\in[0, +\infty),
\label{ds}
\end{equation}
here and below, $h^\#$ denotes, for any measurable function $h$ on
$(0, +\infty){\times}{\bf R}^3{\times}{\bf R}^3, $the following restriction
to characteristics:
$ h^\#(t, {\bf x}, {\bf p})
=h(t, {\bf x}
+t\frac{{\bf p}}{p_0}, {\bf p}).$
It can be also easily proved that  $\tilde{Q}_n(f^n, f^n)
\in L_{loc}^1({\bf R}^3\times{\bf R}^3)$ and hence $f^n$ is a distributional
solution to  (\ref{arbe}).

It is below shown that
$f^n(t, {\bf x}, {\bf p})
{\geq}0$  for almost every $
(t,{\bf x}, {\bf p})\in [0,+\infty)\times
{\bf R}^3{\times}{\bf R}^3.$ To show this,
another mapping $J_n^+$  is first defined as follows: $J_n^+(f)=\max(0,J_n(f))$ for
$\forall f \in {\cal F}^+=\{f:f\in {\cal F} \hbox{ and }
f\mathop{\geq}\limits^{a.e.}0\}.$
 ${\cal F}^+$ is a subset of ${\cal F}.$ Similarly, it  can be easily shown that the
 mapping $J_n^+$ maps ${\cal F}^+$ into itself and is uniformly contractive
 with an inequality similar to (\ref{contractive}),  i.e., for every $g, h\in{\cal F}^+, $
\begin{equation}
\sup\limits_{0{\leq}t<+\infty}e^{-2C_nt}\int\int_{{\bf R}^3\times
{\bf R}^3}|J_n^+(g)(t, {\bf x}, {\bf p})
-J_n^+(h)(t, {\bf x}, {\bf p})|d^3{\bf x}d^3{\bf p}
\nonumber
\end{equation}
\begin{equation}
{\leq}\frac{1}{2}\sup\limits_{0{\leq}t<+\infty}e^{-2C_nt}
\int\int_{{\bf R}^3\times{\bf R}^3}|g
(t, {\bf x}, {\bf p})
-h(t, {\bf x}, {\bf p})|d^3{\bf x}d^3{\bf p}.
\label{contractivep}
\end{equation}
  Then there exists a unique element $\tilde{f}\in {\cal F}^+$
  such that $J_n^+(\tilde{f})=\tilde{f}$ for almost every $(t,{\bf x},{\bf \xi})\in[0,+\infty)\times
  {{\bf R}^3}\times{{\bf R}^3}.$ Thus, if $\tilde{f}=J_n(\tilde{f})$ for almost every $(t,{\bf x},{\bf \xi})\in [0,+\infty)
  \times{{\bf R}^3}\times{{\bf R}^3},$ $\tilde{f}$ is a distributional solution to (\ref{arbe}) through
  $f_0^n$ in the time interval $[0,+\infty) ,$  and $\tilde{f}\mathop{=}\limits^{a.e.}f^n.$
   It will be below shown that $J_n(\tilde{f})\mathop{=}\limits^{a.e.}\tilde{f},$ or equivalently,
   $J_n(\tilde{f})^\#\mathop{=}\limits^{a.e.}\tilde{f}^\#.$ In fact, by (\ref{conop}), it is known that
   $J_n(\tilde{f})^\#$ is absolutely continuous with respect to $t\in [0,+\infty)$ for
   almost every $({\bf x},{\bf \xi})\in {{\bf R}^3}\times{{\bf R}^3}.$ It may be assumed without loss of
    generality  that $J_n(\tilde{f})^\#(t,{\bf x},{\bf \xi})$ is continuous for all $(t,{\bf x},{\bf \xi}).$
To prove that $J_n(\tilde{f})^\#=\tilde{f}^\#,$ it suffices to prove
that $J_n(\tilde{f})^\# \geq 0.$ Assume that
$J_n(\tilde{f})^\#(t_0,{\bf x}_0,{\bf \xi}_0)<0$ for some point
$(t_0,{\bf x}_0,{\bf \xi}_0) \in [0,+\infty)\times{{\bf
R}^3}\times{{\bf R}^3}.$ Note that $J_n(\tilde{f})^\#(0,{\bf
x}_0,{\bf \xi}_0)=f_0^n({\bf x}_0,{\bf \xi}_0)\geq 0.$ Then two
values $t^*$ and $t_1$ can be found in $[0,t_0)$ such that
$J_n(\tilde{f})^\#(t^*,{\bf x}_0,{\bf \xi}_0)=0$ and
$J_n(\tilde{f})^\#(t,{\bf x}_0,{\bf \xi}_0)<0$ for all
$t\in(t^*,t_1],$ so that $ \tilde{f}^\#(t,{\bf x}_0,{\bf \xi}_0)=0$
for all $t\in [t^*,t_1].$ By (\ref{conop}), it can be known that
$0>J_n(\tilde{f})^\#(t,{\bf x}_0,{\bf \xi}_0)\geq
J_n(\tilde{f})^\#(t^*,{\bf x}_0,{\bf \xi}_0)$ for
${\forall}t\in(t*,t_1).$ This is a contradiction with
$J_n(\tilde{f})^\#(t^*,{\bf x}_0,{\bf \xi}_0)=0$. Therefore
$f^n=\tilde{f}=J_n(\tilde{f})\geq 0$ for almost every $ (t,{\bf
x},{\bf \xi})\in [0,+\infty)\times{{\bf R}^3}\times{{\bf R}^3}.$

The properties of $\{f^n(t, {\bf x},
{\bf p})\}^\infty_{n=1}$ are discussed below.  Obviously, $f^n$
 satisfies
\begin{equation}
 0\leq f^n{\in}L^{\infty}{\cap}L^1((0, T){\times}{\bf R}^3{\times}{\bf R}^3)~~({\forall}T<
+{\infty}).
\label{prop1}
\end{equation}
Furthermore,
by using (\ref{ds}), it can be proved
 that
\begin{equation}
 f^n
(t, {\bf x},
{\bf p}){\in}
C([0, +{\infty});L^1({\bf R}^3{\times}{\bf R}^3)).
\label{prop2}
\end{equation}

For convenience, $\tilde{Q}_n^+, \tilde{L}_n$ and $\tilde{Q}_n^-$ are, respectively,  denoted by
\begin{equation}
\tilde{Q}_n^+(\varphi,\varphi)=(1+\frac{1}{n}\int_{{\bf R}^3}|\varphi | d^3{\bf p})^{-1}\frac{1}{p_0}\int_{{\bf R}^3}\frac{d^3 {\bf p}_1}{p_{1 0}}
\int_{{S}^2}d\Omega \varphi({\bf p}^\prime)\varphi({\bf p}_1^\prime)B_n(g, \theta),
\label{arbekp}
\end{equation}
\begin{equation}
\tilde{L}_n(\varphi)=(1+\frac{1}{n}\int_{{\bf R}^3}|\varphi | d^3{\bf p})^{-1}\frac{1}{p_0}\int_{{\bf R}^3}\frac{d^3 {\bf p}_1}{p_{1 0}}
\int_{{S}^2}d\Omega \varphi({\bf p}_1)B_n(g, \theta).
\label{arbekl}
\end{equation}
\begin{equation}
\tilde{Q}_n^-(\varphi,\varphi)=\varphi({\bf p})\tilde{L}_n(\varphi),
\label{arbekm}
\end{equation}
Obviously,
\begin{equation}
\tilde{Q}_n(\varphi,\varphi)=\tilde{Q}_n^+(\varphi,\varphi)-\tilde{Q}_n^-(\varphi,\varphi).
\label{arbekpm}
\end{equation}
Similarly,  (\ref{arbek}) is equivalent to the form
\begin{equation}
Q_n(\varphi, \varphi)=\frac {1}{2}\mathop{\int\int\int}
\limits_{{\bf R}^3{\times}{\bf R}^3{\times}{\bf R}^3}
\frac{W_n({\bf p}, {\bf p}_1;{\bf p}^\prime, {\bf p}_1^\prime)}
{p_0p_{10}p_0^{\prime}p_{10}^\prime}
[\varphi^{\prime}\varphi_1^\prime-\varphi\varphi_1]d^3{\bf p}_1d^3{\bf p}^{\prime}d^3{\bf p}_1^\prime
\label{arbektr}
\end{equation}
with
\begin{equation}
 W_n({\bf p}, {\bf p}_1;{\bf p}^\prime, {\bf p}_1^\prime)
=s\sigma_n(g, \theta){\delta}^{(3)}({\bf p}+{\bf p}_1-{\bf p}^\prime-{\bf p}_1^\prime)
\delta(p_0+p_{10}-p_0^\prime-p_{10}^\prime).
\label{arbetr}
\end{equation}
If $\tilde{Q}^\pm_n(\varphi, \varphi)\psi({\bf p})\in{L}^1({\bf R}^3)
$ for every given $\psi({\bf p})\in{L}^\infty
({\bf R}^3)$
 and $\varphi
({\bf p})\in{L}^1({\bf R}^3), $ then
   it can be deduced by using (\ref{arbetr}) that
\begin{equation}
\int_{{\bf R}^3}\psi({\bf p})
\tilde{Q}_n(\varphi, \varphi)d^3p
=\frac {1}{4}(1+\int_{{\bf R}^3}\varphi d^3{\bf p})^{-1}
\iiint_{{\bf R}^3\times{\bf R}^3\times{S}^2}\frac{d^3 {\bf p}_1d^3 {\bf p}d\Omega }
{p_{1 0}p_0} B_n(g, \theta)
\nonumber
\end{equation}
\begin{equation}
\cdot
[\varphi^{\prime}\varphi_1^\prime-\varphi\varphi_1][\psi({\bf p})+\psi(
{\bf p}_1)
-\psi({\bf p}^\prime)-\psi(
{\bf p}_1^\prime)] .
\label{arbeken}
\end{equation}

By (\ref{arbekp}), (\ref{arbekl}) and (\ref{arbekm}),
it is obvious to see that
\begin{equation}
\tilde{Q}^+_n(f^n, f^n), \tilde{Q}^-_n(f^n, f^n)
{\in}L^1_{loc}((0, +{\infty}){\times}{\bf R}^3{\times}{\bf R}^3).
\label{prop3}
\end{equation}
Then, by (\ref{dinvariant3}), (\ref{rbeei}), (\ref{arbeken})
and Gronwall's inequality, it can be found that
\begin{equation}
 \sup\limits_{0\leq t\leq T}{{\int}{\int}}_{{\bf R}^3{\times}{\bf R}^3}
 f^n(1+|{\bf x}|^2+p_0
+|{\ln}f^n|)d^3{\bf x}d^3{\bf p}{\leq}C_T.
\label{prop4}
\end{equation}
It also follows by (\ref{rbeei}) that
\begin{equation}
 \frac{1}{4}{\int}^{+\infty}_0{\int}_{{\bf R}^3} \left\{(1+{\int}_{{\bf R}^3} f^n
d^3 {\bf p})^{-1}\iiint_{{\bf R}^3\times{\bf R}^3\times{S}^2}\frac{d^3 {\bf p}_1d^3 {\bf p}d\Omega }
{p_{1 0}p_0} \right.B_n(g, \theta)
\nonumber
\end{equation}
\begin{equation}
\cdot\left.(f^{n{\prime}}f^{n{\prime}}_1-f^nf^n_1)
{\ln}\left(\frac{f^{n{\prime}}f^{n{\prime}}_1}
{f^nf^n_1}\right)\right\}d{\sigma}d^3 {\bf x}
 {\leq}C_T.
\label{prop5}
\end{equation}
In (\ref{prop4}) and (\ref{prop5}) $C_T$ is a positive constant only dependent of
 $f_0$ and $T$ except of $n.$
By using  necessary and sufficient conditions of a weakly
compact set in the $L^1$-space (see \cite{ds58}, Page 347, IV.13.54),
it can be deduced from (\ref{prop4}) that
$\{f^n\}^{\infty}_{n=1}$ is
  weakly compact in $L^1((0, T){\times}
{\bf R}^3{\times}{\bf R}^3)$ $({\forall}T<+{\infty})$ and so it may be assumed
without loss of generality that $ f^n$ converges weakly in
$L^1((0, T){\times}{\bf R}^3{\times}{\bf R}^3)$ to
$f{\in}L^1_{loc}([0, +{\infty}){\times}{\bf R}^3{\times}{\bf R}^3)$
as $n{\rightarrow}+{\infty}$
 for all $T\in (0,+{\infty}).$  Obviously, $f\geq 0$ and
 $f|_{t=0}=f_0({\bf x},
{\bf p})$ for almost every $({\bf x}, {\bf p}){\in}{\bf R}^3{\times}
{\bf R}^3.$  It is also known that (\ref{prop4}) yields (\ref{sc4}).

It can be also claimed that for ${\forall}T, R\in (0,+{\infty}),$ $\{\tilde{Q}_n^\pm(f^n,f^n)/(1+f^n)\}^{\infty}_{n=1} $
are weakly compact subsets of
$L^1((0, T){\times}{\bf R}^3{\times}B_R).$

Once this claim is proven, it follows that  $f$ is a global mild
solution to RBE (\ref{rbe}) satisfying
(\ref{sc1}), (\ref{sc2}), (\ref{sc3}) and (\ref{sc4}), by analyzing step by step
 the subsolutions and supersolutions of RBE (\ref{rbe}) with a similar device to that
 given by DiPerna and Lions \cite{dl}.
Our analysis not only allows for the relations among
three different types of solutions to RBE (\ref{rbe}) (see \cite{j98a}) but also requires
the momentum-averaged compactness of the transport operator
of RBE (\ref{rbe}) (see \cite{gl} or \cite{j97}).

It remains to show this claim. In the case of
$\tilde{Q}_{n}^-,$ to prove this, it suffices to
 prove that  $\tilde{L}_{n}(f^n)$
belongs to some weakly
compact subset of $ L^1((0,T)\times{\bf R^3}\times{B}_R).$
Denote $\tilde{L}_{nk}~(k=1,2,3,\cdots) $ by
\begin{equation}
\tilde{L}_{nk}(f^n)=(1+\frac{1}{n}\int_{{\bf R}^3}|f^n| d^3{\bf p})^{-1}\frac{1}{p_0}\int_{{\bf R}^3}\frac{d^3 {\bf p}_1}{p_{1 0}}
\int_{{S}^2}d\Omega f^n({\bf p}_1)B_n(g, \theta)1_{|{\bf p}_1|\leq k}.
\label{rbeklk}
\end{equation}
Then put $\Psi(y)=y(\ln{y})^+$ and define
$\bar{b}_k:=\bar{b}_k(t,{\bf x},{\bf p})$ by
\begin{equation}
\bar{b}_k=(1+\frac{1}{n}\int_{{\bf R}^3}|f^n| d^3{\bf p})^{-1}\frac{1}{p_0}\int_{{\bf R}^3}\frac{d^3 {\bf p}_1}{p_{1 0}}
\int_{{S}^2}d\Omega B_n(g, \theta)1_{|{\bf p}_1|\leq k}.
\label{rbelk}
\end{equation}
Thus
\begin{equation}
\Psi(\tilde{L}_{nk}(f^n)){\leq}\bar{b}_k\Psi(\frac{\tilde{L}_{nk}(f^n)}
{\bar{b}_k})+(\ln\bar{b}_k)\tilde{L}_{nk}(f^n).
\label{pineq1}
\end{equation}
Take $\alpha_{n}=|f^n|_{L^\infty}+1 $ and
$v=\bar{b}_k^{-1}\tilde{L}_{nk}(f^n).$
Since $\Psi(y) $ is convex, there exists a positive constant
$\beta_{n}$ which
depends only on $t,{\bf x},
{\bf p}$ and ${n},$
such that, for $0\leq{u}<v<w<\alpha_{n},$
$$0\leq\frac{\Psi(u)-\Psi(v)}{u-v}\leq\beta_{n}
\leq\frac{\Psi(v)-\Psi(w)}{v-w},
$$
which gives $$
\Psi(u)\geq\Psi(v)+\beta_{n}(u-v)
~~(\forall u\in[0,\alpha_{n}]),$$
in particular,
$$\Psi(f_1^{n})\geq\Psi(v)+\beta_{n}
(f_1^{n}-v).$$
It follows that
\begin{equation}
\Psi(\frac{\tilde{L}_{nk}(f^n)}{\bar{b}_k})\leq
\frac {1}{\bar{b}_k}
\int\int_{{\bf R^3{\times}}S^2}\frac{B_n(g,\theta)}
{p_0p_{10}}\Psi(f_1^{n})
1_{|{\bf p}_1|{\leq}k}
d^3{\bf p}_1d\Omega.
\label{pineq2}
\end{equation}
By (\ref{pineq1}) and (\ref{pineq2}), it can be concluded that
\begin{eqnarray}
\int\int_{{\bf R^3}\times{B}_R}\Psi(\tilde{L}_{nk}(f^n))d^3{\bf x}
d^3{\bf p} \hfill \hspace*{6cm}
\nonumber \\ \leq
\int\int_{{\bf R^3}\times{B}_R}\left\{
\int\int_{{\bf R^3{\times}}S^2}\frac{B_n(g,\theta)}
{p_0p_{10}}\Psi(f_1^{n})
1_{|{\bf p}_1|{\leq}k}
d^3{\bf p}_1d\Omega\right\}d^3{\bf x}d^3{\bf p}
\nonumber \\+\int\int_{{\bf R^3}\times{B}_R}
(\ln\bar{b}_k)\tilde{L}_{nk}(f^n)d^3{\bf x}d^3{\bf p}.
\label{pineq3}
\end{eqnarray}
It can be known from (\ref{rbeb1}) that $0\leq\bar{b}_k\leq{b}_{Rk}$ for all $
(t,{\bf x},{\bf p})\in(0,+\infty)
\times{\bf R^3}\times{B}_R$ and
\begin{eqnarray}
\int\int_{{\bf R^3{\times}}S^2}\frac{B_n(g,\theta)}
{p_0p_{10}}1_{|{\bf p}_1|{\leq}k}1_{|{\bf p}|{\leq}R}d^3{\bf p}d\Omega
\hfill \hspace*{3cm}
\nonumber \\
{\leq}\int\int_{{\bf R^3{\times}}S^2}\frac{B(g,\theta)}
{p_0p_{10}}1_{|{\bf p}_1|{\leq}k}
1_{|{\bf p}|{\leq}R}d^3{\bf p}d\Omega{\leq}b_{Rk},
\label{kineq}
\end{eqnarray}
where $b_{Rk}$ is a positive constant which depends only on $R$ and $k,$
thus it can be shown from (\ref{rbeklk}) and (\ref{pineq3}) that
$$\int\int_{{\bf R^3}\times{B}_R}\Psi(\tilde{L}_{nk}(f^n))d^3{\bf x}d^3{\bf p}
$$$${\leq}b_{Rk}\int\int_{{\bf R^3}\times{\bf R^3}}
\Psi(f^n)d^3{\bf x}d^3{\bf p}+
b_{Rk}|\ln{b}_{Rk}|\int\int_{{\bf R^3}\times{\bf R^3}}
f^nd^3{\bf x}d^3{\bf p}.$$
Hence, by (\ref{prop4}), $\{\Psi(\tilde{L}_{nk}(f^n))\}_{n=1}^\infty$ is bounded in
$L^\infty((0,T);L^1({\bf R^3}\times{B}_R))$ for every given $k.$
It can be also
known from (\ref{prop4}), (\ref{rbeklk}) and (\ref{kineq}) that
\begin{equation}
\sup\limits_{n\geq 1}\int\int_{{\bf R^3}\times{B}_R}\tilde{L}_{nk}
(f^n)(1+|{\bf x}|^2+p_0)
d^3{\bf x}d^3{\bf p}<C_{Rk}
\label{hand01}
\end{equation}
where $C_{Rk}$ is a positive constant which only depends on $R$ and $k.$
It follows that $\tilde{L}_{nk}(f^n)$ belongs to a
weakly compact subset of $L^1((0,T)\times{\bf R^3}\times{B}_R)$ for any given
 $R$ and $k$ in $(0,+\infty).$

On the other hand,  it can be shown that as $k\to+\infty,$
\begin{equation}
\sup\limits_{t\in[0,T]}\sup\limits_{{n}\geq 1}|\tilde{L}_{nk}(f^n)
-\tilde{L}_{n}(f^n)|
_{L^1({\bf R^3}\times{B}_R)}\longrightarrow{0}.
\label{hand02}
\end{equation}
Indeed, it can be known from  (\ref{rbeb2})  that,
for every small $\sigma>0,$ there exists $k_0>0,$
such that, as $|{\bf p}_1|>k_0,$
\begin{equation}
 \int\int_{B_R\times{S}^2}
\frac{B(g,\theta)}{p_0p_{10}}d^3{\bf p}d\Omega<{\sigma}
p_{10}.
\label{pineq4}
\end{equation}
By using (\ref{rbeklk}) and (\ref{pineq4}),  it can be shown that, as $k>k_0,$
$$\sup\limits_{t\in[0,T]}\sup\limits_{{n}\geq 1}|\tilde{L}_{nk}(f^n)
-\tilde{L}_{n}(f^n)|
_{L^1({\bf R^3}\times{B}_R)}$$
$$\leq\sup\limits_{t\in[0,T]}\sup\limits_{{n}\geq 1}
\int\int_{{\bf R^3}\times{\bf R^3}}
\left\{\left[
\int\int_{B_R\times{S}^2}
\frac{B(g,\theta)}{p_0p_{10}}d^3pd\Omega\right]\cdot
1_{|{\bf p}_1|>k}{f}_1^{n}
\right\}d^3{\bf x}d^3{\bf p}_1 $$$$
\leq\sigma \cdot\sup\limits_{t\in[0,T]}\sup\limits_{{n}\geq 1}
\int\int_{{\bf R^3}\times{\bf R^3}}p_{10}f_1^{n}d^3{\bf x}d^3{\bf p}_1.$$
This  leads  easily to (\ref{hand02}) by using (\ref{prop4}).

Therefore, by (\ref{hand01}) and (\ref{hand02}),
it can be easily deduced that $\{\tilde{L}_{n}(f^n)\}$
is weakly compact in $L^1((0,T)$ $\times{\bf R^3}\times{B}_R).$

In the case of $\tilde{Q}_{n}^+,$
the following inequality is used: $$\tilde{Q}_{n}^\pm(f^n,f^n){\leq}
K\tilde{Q}_{n}^\mp(f^n,f^n)
+\frac{\tilde{e}^{n}}{\ln{K}}$$ for every $K>1,$
where $\tilde{e}^{n}$ is defined by
$$\tilde{e}^{n}=\iiint_{{\bf R}^3\times{\bf R}^3\times{S}^2}\frac{d^3 {\bf p}_1d^3 {\bf p}d\Omega }
{p_{1 0}p_0}\frac{B_n(g, \theta)
(f^{n\prime}f_1^{n\prime}
-f^nf_1^{n})}{1+\frac{1}{n}\int_{{\bf R}^3}|f^n| d^3{\bf p}}\ln\left(\frac{
f^{{n}\prime}f_1^{{n}\prime}}
{f^nf_1^{n}}\right),
 $$
and thus, by (\ref{prop5}) and the above result in the case of $\tilde{Q}_{n}^-,$ it can be shown that
$\frac{\tilde{Q}_{n}^+(f^n,f^n)}{1+f^n}$
belongs to some weakly compact subset of $L^1((0,T)\times{\bf R^3}
\times{B}_R)$
 for any given $R$ and $T$ in $(0,+\infty).$

The proof of this claim is finished and the theorem holds.

%\noindent
\renewcommand{\thetheorem}{\arabic{theorem}}
\begin{remark}
The proof of the existence and uniqueness of global solution
to the Cauchy problem (\ref{arbe})  is simpler than that given in \cite{j98a}.
\end{remark}
\begin{remark}
By using the device of Dudy\'{n}ski and Ekiel-Je\.{z}ewska  \cite{de92},
 the causality of RBE (\ref{rbe}) (see \cite{de85}\cite{de85e}) can be directly applied into
the further proof of the existence of global solution to the Cauchy
problem for RBE (\ref{rbe}) in some relativistic hard interaction
cases with the finite initial physically natural bounds excluding
the finite initial ``inertia'', i.e., with the initial data
$f_0({\bf x},{\bf p})$ satisfying
\begin{equation}
f_0\mathop{\geq}\limits^{a.e.}0,
\iint_{{\bf R}^3\times{\bf R}^3}f_0(1+p_0+|\ln f_0|)d^3{\bf x}d^3{\bf p}<\infty.
\label{rbec3}
\end{equation}
Exactly speaking,
under the assumptions of (\ref{rbeb1}), (\ref{rbeb2}) and (\ref{rbec3}),
RBE (\ref{rbe}) has a mild
or equivalently a renormalized solution $f$
through initial data $f_0,$
satisfying (\ref{sc1}),  (\ref{sc2}), (\ref{sc3}) and
\begin{equation}
\sup\limits_{t\geq 0}\iint_{{\bf R}^3{\times}{\bf R}^3}f(1+p_0
+{\ln}f)d^3{\bf x}d^3{\bf p}<+\infty. \label{sc5}\end{equation} The
first author of this paper will give another paper to show a detail
proof of this result.
\end{remark}

{\small\noindent{\sc\bf Acknowledgement.}~
This work was supported by grants of NSFC 10271121 and
joint grants of NSFC 10511120278/10611120371 and RFBR 04-02-39026.
This work was also sponsored by SRF for ROCS, SEM. 
We thank Professor Hisao Fujita Yashima very much for his finding 
two typos in the Chinese Edition of this paper: 
one is the definition of the variable $g$ and the other the sign after the first term 
on the right side of the equality in (\ref{enineq}). 
We would like to thank the referee of this paper
for his/her valuable comments on this work.}

\end{document}